\documentclass[twocolumn,aps]{revtex4}
\usepackage{graphicx}
\usepackage{dcolumn}
\usepackage{bm}
\usepackage{amssymb}
\usepackage{amsmath}

\begin{document}

\title{Effects of Space Charge, Dopants, and Strain Fields on Surfaces 
and Grain Boundaries in YBCO Compounds}

\author{Haibin Su}
\email{hbsu@wag.caltech.edu}
\altaffiliation{Present address: Beckman Institute 139-74, California 
Institute of Technology, Pasadena, CA 91125}
\affiliation{Department of Materials Science and Engineering, SUNY at Stony Brook, Stony Brook, NY 11794}
\author{David O. Welch}
\email{dwelch@bnl.gov}
\affiliation{Department of Materials Science, Brookhaven National Laboratory, Upton, NY 11973}

\date{\today}

\begin{abstract}
\noindent Statistical thermodynamical and kinetically-limited models are applied to
study the origin and evolution of space charges and band-bending effects at low angle
[001] tilt grain boundaries in YBa$_2$Cu$_3$O$_7$ and the effects of Ca 
doping upon them. Atomistic simulations, using shell models of 
interatomic forces, are used to
calculate the energetics of various relevant point defects. The
intrinsic space charge profiles at ideal surfaces are calculated for two limits of oxygen contents, i.e.
YBa$_2$Cu$_3$O$_6$ and YBa$_2$Cu$_3$O$_7$. At one limit, O$_6$, the system is an
insulator, while at O$_7$, a metal. This is analogous to the intrinsic and doping
cases of semiconductors. The site selections for doping calcium and creating holes are
also investigated by calculating the heat of solution. In a continuum treatment, the
volume of formation of doping calcium at Y-sites is computed. It is then applied to
study the segregation of calcium ions to grain boundaries in the Y-123
compound. The influences of the segregation of calcium ions on space charge profiles
are finally studied to provide one guide for understanding the improvement of
transport properties by doping calcium at grain boundaries in Y-123 compound.

\end{abstract}

\maketitle

\section{Introduction}

The study of grain boundaries of YBCO is very important, because they act
as strong barriers to current flow in YBCO \cite{mannhart02}. This behavior of high
temperature superconductors is in sharp contrast to metallic low temperature
superconductors, in which grain boundaries are not only transparent to current, but
also significantly contribute to flux pinning. The formation of such weak links in
high temperature superconductors is due to the low carrier density and strong
dependence of $T_c$ on hole concentration. There are many experiments showing that
the critical current density through grain boundaries is very sensitive to the
orientation of the adjacent crystallines \cite{Dimos88, Dimos90}. Generally, a large
misorientation angle will reduce $J_c$ significantly. As we know, there are quite
rich phenomena occurring in grain boundaries: variations in the chemical
stoichiometry, dissociation of grain boundary dislocations, impurity segregation,
etc.  For instance, recently Kung et al \cite{kung01} report that small angle [001]
tilt grain boundaries in YBCO with (110) planes exhibit partial grain boundary
dislocations separated by stacking faults. The dissociated grain boundary structures
have twice the number of grain boundary dislocations and shorter interdislocation core
channel widths than Frank's geometry rule predicts (in Frank's rule the separation
between dislocations of small angle $\theta$ boundaries is proportional to $\rm {b
\over \theta}$). This requires further development of the above model of supercurrent
flow through arrays of small angle grain boundaries. Furthermore, the situation gets
more complicated because grain boundaries can also act as sinks and sources for
vacancies. In ionic solids this can give rise to charging effects. The existence of
space charge regions near free surfaces in an ionic solid was first postulated by
Frenkel \cite{Frenkel46}. In an ionic solid, the vacancy concentrations at surfaces and grain
boundaries are determined by the individual free energy of formation. The cation and
anion vacancies usually have different formation energies. However, the vacancy
concentrations in the bulk are determined by the condition of charge neutrality and
usually they are different from those at the grain boundaries 
where strict charge neutrality conditions are relaxed. Thus, the vacancy
concentration changes as a function of distance from the grain boundaries. Usually,
vacancies in ionic solids are charged defects. Spatial distributions of these charged
defects lead to a potential difference between the bulk and free surfaces and grain
boundaries \cite{Lehovec53, Kliewer65, Yan83a, Yan83b}. There are two kinds of space
charges. Here we define a space charge arising from differences in point defect
formation energies as ``intrinsic space charge" to contrast with those due to
segregated dopants and grain boundary nonstoichiometries.

The topic of space charges at interfaces arising from point defects has
not been investigated so far for
cuprates. Since holes carry a charge, it is expected that a space
charge will produce a significant effect on hole distributions
around grain boundaries for the metallic phase of
$\rm YBa_2Cu_3O_{7-\delta}$.
Actually the holes can segregate around charged boundaries even if not
metallic. These can be localized hole distributions.
The generic phase diagram of the cuprates shows a wide variety of different behaviour at different
temperatures and levels of doping \cite{Zhang93, Tallon95}. At zero doping
the cuprates are all insulators, and below a few hundred kelvin they are
also antiferromagnets (i.e. the electron spins on neighbouring copper
ions point in opposite directions). However,
when the doping is increased above a critical value (about $5\%$,
although this varies from compound to compound), the
antiferromagnetic state disappears and we enter the so-called
underdoped region. As the doping is increased further,
superconducting region is reached. For $\rm YBa_2Cu_3O_{7-\delta}$,
the doping is closely related to the oxygen content. Here we examine two
limits: $\delta = 0, 1$. When $\delta = 1$, this is an insulator,
which can be approximately treated as an ionic solid. However, when
$\delta = 0$, the phase is superconducting, and has mobile
holes. The existence of holes can add more screening than in the
insulator case. This is quite similar to the two limits of
semiconductors: point defect limit and electron/hole limit.
As the oxygen content changes from O$_7$ to O$_6$, the c-axis expands
\cite{cava87}. The lattice strain has a large effect on point defects \cite{hbsu-unpub}. Besides, the valence of chain 
copper ions decreases from +2 to +1 as the oxygen content is reduced. Both factors lead to a dramatic change of point 
defect formation energies in
123 compounds. Consequently, this will affect the intrinsic space charge.

It has been known since the early days of YBCO that doping with calcium is one way of
adding more holes to the material. Divalent calcium is roughly the same size as
trivalent yttrium and so readily replaces it. If a deficiency of holes causes grain
boundaries to have weak coupling, then the critical current might be improved by
doping calcium into grain boundaries. Note that if calcium is doped inside the bulk
as well, the bulk $T_c$ will be lowered. Therefore, doping calcium at the grain
boundaries can both repair transport properties of grain boundaries and keep the high
$T_c$ value of the bulk. Indeed, Hammerl et al \cite{Hammerl00} recently developed
one beautiful process to increase the critical current density above $10^5 (A / cm^{2})$ at temperatures near the boiling 
point of liquid nitrogen by doping Ca into only high angle grain boundaries of YBCO. So, the current key issue is to 
improve transport properties near low-angle [001] tilt grain boundaries, which are the main
types of defects in polycrystalline YBCO wires and tapes
\cite{guth01,durrell03,foltyn03}. For small angle grain boundaries, Gurevich et al
\cite{Gurevich98} argue that the critical current density dependence is mostly
determined by the decrease of the current-carrying cross section by insulating
dislocation cores and by progressive local suppression of the superconducting order
parameter near grain boundaries as $\theta$ increases. However, space charge
effects due to Schottky disorder are not considered in their model, although there
exist clear space charge profiles in cuprate superconductors \cite{zhu-spch}. It is thus
desirable to make some theoretical inputs for this topic, which can help to provide useful guides
to find practical ways to improve the transport properties of grain boundaries in
YBCO.

In this paper, first we construct a statistical thermodynamic model to study the origin and evolution of 
space charges due to the spectrum of point defect formation energies as a function of the
variation of oxygen content in YBa$_2$Cu$_3$O$_{7-\delta}$. Then we develop one kinetically-limited model, including 
effects of doping and solute segregation, to study the process of Ca doping upon low angle tilt grain boundaries in YBCO 
and consequential influences on the space charge profile.

\section{Space Charge Profile for the Intrinsic and Doping Cases for 123
compounds}

The simplest imperfection in a crystal lattice is a lattice vacancy, which is a missing atom or ion. 
Electrically neutral and structure-preserving groups of vacancies, known as Schottky defects, are
required in compounds. No matter by what means a Schottky defect is made, it is
necessary to expend a certain amount of work per atom taken to the surface.
Therefore, the energy of the crystal is increased. At a finite temperature, 
the average energy of formation of the group of Schottky defects determines 
the concentration of such point defects at equilibrium. For a
multicomponent crystal, the difference of the energy of formation of various defects
determines the nature of space charge near interfaces \cite{Frenkel46}. We calculated
the point defect formation energies for both YBa$_2$Cu$_3$O$_6$ and
YBa$_2$Cu$_3$O$_7$ by the Mott-Littleton approach \cite{mott38,mott48} which has been
implemented in GULP codes \cite{gulp}. The pair potentials of the shell model used in
this study were developed by Baetzold for Y-123 for studying ionic and electronic
(polaron) defects \cite{2baetzold88}. The calculated point defect formation energies
are collected in Table (\ref{tab:PointDefect}). The data for YBa$_2$Cu$_3$O$_7$ agree
nicely with the previous results \cite{2baetzold88} using HADES III codes \cite{hades74}.
It can be seen from Table (I) that there is a wide spectrum in the calculated formation energies of individual vacancies 
in the Schottky group for both YBa$_2$Cu$_3$O$_6$ and YBa$_2$Cu$_3$O$_7$. Thus, in principle there should be space 
charges near free surfaces at equilibrium or quenched in from high temperatures where ionic mobility is high enough to 
maintain equilibrium. A similar phenomenon is expected to occur near internal interfaces such as grain boundaries, but 
the situation is more complex than for free surfaces for at least two reasons. First of all, the spectrum of dopant 
formation energies may be different for the process of removing an ion from a site in the bulk 
crystal and placing it at a site in the grain boundary. Of course, at equilibrium the chemical potential of defects is 
the same throughout the crystal, but the local variations in concentration will be affected by different energies for 
excess species (i.e. the ions removed to form the vacancies and placed at surface or boundary sites) in different types 
of interfaces. The second complicating factor is that the local stoichiometry in the structural units (or individual 
dislocation cores) making up the boundary may be significantly different from that of the bulk 
in complex ionic crystals. Advanced transmission electron microscopy has been used recently to show that this is the 
case for grain boundaries 
in SrTiO$_3$ \cite{pennycook94} and for twin boundaries in BaTiO$_3$ \cite{jia04}. It is not entirely clear how such 
local chemistry at boundaries in complex oxides 
will affect the intrinsic space charges arising from the spectrum of defect formation energies in equilibrium groupings, 
e.g. Schottky defects. Therefore, we will restrict our detailed calculations of intrinsic space charge to free surfaces, 
but we expect the behavior at grain boundaries to be at least qualitatively similar.

\begin{table}
\begin{center}
\caption{Point defect formation energy (eV) in YBa$_2$Cu$_3$O$_6$ and
YBa$_2$Cu$_3$O$_7$. This defect formation energy refers to removing one ion from its
lattice site and putting it at ledge-corner site on the surface of the crystal. }
\label{tab:PointDefect}
\begin{tabular}{rrr}
\hline
&  $\rm YBa_2Cu_3O_6$ &  $\rm YBa_2Cu_3O_7$ \\ \hline
      E(Y) &       14.42 &     12.71 \\
     E(Ba) &       4.06 &       4.25 \\
   E(Cu-chain) &   0.96 &       4.23 \\
   E(Cu-plane) &   6.92 &       5.68 \\
E(O-plane) &       3.47 &       3.56 \\
E(O-chain) &            &       4.22 \\
 E(O-apex) &       6.85 &       4.48 \\ \hline
\end{tabular}
\end{center}
\end{table}

\subsection{Intrinsic Space Charge at Surfaces}

\subsubsection{Insulator Case: YBa$_2$Cu$_3$O$_6$}

For the sake of simplicity, we assume the crystal has free surfaces at x=0 and x=2L and is of infinite extent in
the y and z directions in studying intrinsic space charge problems, i.e. only containing free surfaces. 
As discussed above, the case of
grain boundaries is more complex, but the calculations for free surfaces will provide some qualitative guide to
understanding of intrinsic space charges, i.e. those due to equilibrium Schottky defects at other interfaces too. The 
free energy per unit area of a slab-like
disordered crystal of the thickness 2L is given by

\begin{widetext}
\begin{equation} \rm
F = \int [n_YE_Y + n_{Ba}E_{Ba} + n_{Cu_c}E_{Cu_c} + n_{Cu_p}E_{Cu_p} +
n_{O_p}E_{O_p} + n_{O_a}E_{O_a} + {1 \over 2} \rho(x)\Phi(x)]dx -
TS_{conf};
\end{equation}
\end{widetext}

\noindent where $\Phi (x)$ is the electrostatic potential, $\rho (x)$ is the local charge density, 
and $\rm n_{Y}, n_{Ba}, n_{Cu_c}, n_{Cu_p}, n_{O_p}$, and
$\rm n_{O_a}$ stand for the densities (per unit volume) of vacancies at yttrium, barium,
chain copper, plane copper, plane oxygen, and apex oxygen sites, respectively. Note that the
configurational entropy density $\rm S_{conf}$ is calculated as

\begin{widetext}
\begin{equation} \rm
{\delta S_{conf} \over k} = \delta n_Y \ln {N \over n_Y} + \delta
n_{Ba} \ln {2N \over n_{Ba}} +  \delta n_{Cu_c} \ln {N \over
n_{Cu_c}} +  \delta n_{Cu_p} \ln {2N \over n_{Cu_p}} + \delta
n_{O_p} \ln {4N \over n_{O_p}} + \delta n_{O_a} \ln {2N \over
n_{O_a}},
\end{equation}
\end{widetext}

\noindent where N is the number of formula units per unit volume. This arises because we have 
neglected effects of clustering and association (e.g. the formation of di-vacancies etc.) among the defects.
The charge density $\rm \rho (x)$ is given by

\begin{equation} \rm
\label{eq:chargeO6} \rho(x) = e ( 2n_{O_p}+ 2n_{O_a} - 3 n_Y -2
n_{Ba} - n_{Cu_c} - 2n_{Cu_p}).
\end{equation}

From $\delta$F = 0, the spatially-varying equilibrium concentrations are

\begin{equation}
\label{eq:concenO6}
\left\{
\begin{array}{cc}
\rm    n_Y    =  N exp(- {E_Y-3e \Phi \over kT}), \\
\rm    n_{Ba} = 2N exp(- {E_{Ba}- 2e \Phi \over kT}), \\
\rm    n_{Cu_c} = N exp(- {E_{Cu_c}- e \Phi \over kT}), \\
\rm    n_{Cu_p} = 2N exp(- {E_{Cu_p}- 2e \Phi \over kT}), \\
\rm    n_{O_p} = 4N exp(- {E_{O_p} + 2e \Phi \over kT}), \\
\rm    n_{O_a} = 2N exp(- {E_{O_a} + 2e \Phi \over kT}). \\
\end{array}
\right.
\end{equation}

The electrostatic potential $\rm \Phi(x)$ is obtained from Poisson's
equation, which can be written Using equations (\ref{eq:chargeO6}, \ref{eq:concenO6}) as

\begin{widetext}
\begin{eqnarray}
\label{eq:poissonO6} \rm
\nabla^2 \Phi(x) & = & {-4\pi \over \epsilon} \rho(x) 
\nonumber \\
& = & {-4 \pi e N \over \epsilon} [ 2*4*exp(- {E_{O_p}+
2e\Phi \over kT }) +  2*2*exp(- {E_{O_a}+ 2e\Phi \over kT })
- 3*1*exp(-{E_Y-3e \Phi \over kT})     
\nonumber \\
& - & 2*2*exp(-{E_{Ba}-2e \Phi \over kT })-1*1*exp(-{E_{Cu_c} -
e \Phi \over kT})-2*2*exp(-{E_{Cu_p}-2e \Phi \over kT}) ];
\end{eqnarray}
\end{widetext}

\noindent where $\epsilon$ is the static relative permittivity 
($\epsilon$ = 5.21), which is obtained by atomistic simulations using our shell
model\cite{su-thesis} with pair potential parameters from 
Baetzold\cite{2baetzold88}. We define the electrostatic potential to be 
zero at the surface, and the bulk potential $\Phi_\infty$ is defined as 
the potential for which $\rho = 0$. Therefore, letting the 
right-hand-side(RHS) of Eq. (\ref{eq:poissonO6}) equal zero, the bulk 
potential $\Phi_\infty$ can be obtained, and the defects' concentrations 
in bulk can be calculated. We also define the Debye length $\kappa$ 
\cite{Frenkel46,Lehovec53,Kliewer65} as

\begin{equation}
\kappa = ( { \epsilon kT \over 8\pi e^2 \sum_i  (n_iq_i^2) } )^ {
1 \over 2 },
\end{equation}

\noindent where $\rm n_i$ is the density of vacancies on sites of type i, and $\rm
q_i$ is the charge of the vacancy.

\begin{table}
\begin{center}
\caption{Bulk Potential and Debye Length of
$\rm YBa_2Cu_3O_6$.}\label{tab:O6}
\begin{tabular}{rrr}
\hline
     T (K) &   $\rm {e\Phi_{\infty}\over kT}$ &   $\kappa$ (m) \\ \hline
       350 &   -27.42 &   44.34 \\
       400 &   -23.81 &   0.93 \\
       600 &   -15.61 &   1.83x10$^{-4}$ \\
      1000 &    -9.28 &   2.49x10$^{-7}$ \\  \hline
\end{tabular}
\end{center}
\end{table}

If we scale the original length r by the Debye length $\kappa$, and define
one new variable z as follows,

\begin{eqnarray} \rm
s & = & {x \over \kappa }, \nonumber \\
z & = & {(e\Phi - e\Phi_\infty) \over kT },
\end{eqnarray}

\noindent equation (\ref{eq:poissonO6}) can be rewritten as

\begin{widetext}
\begin{eqnarray}
\label{eq:o6o6} \rm
{ (d^2z) \over (ds^2) } & = & \rm -{1 \over 2}[2*4*exp(- {E_{O_p}+
2e\Phi_\infty \over kT })\exp(-2z) + 2*2*exp(- {E_{O_a}+
2e\Phi_\infty \over kT })\exp(-2z)  \nonumber \\
\rm
& - & 3*1*exp(-{E_Y-3e \Phi_\infty \over kT})\exp(3z)
-2*2*exp(-{E_{Ba}-2e \Phi_\infty \over kT})\exp(2z)  \nonumber \\
\rm
& - & 1*1*exp(-{E_{Cu_c}- e \Phi_\infty \over kT})\exp(z) -
2*2*exp(-{E_{Cu_p}-2e \Phi_\infty \over kT})\exp(2z) ].
\end{eqnarray}
\end{widetext}

\begin{figure}
\includegraphics [scale=0.4] {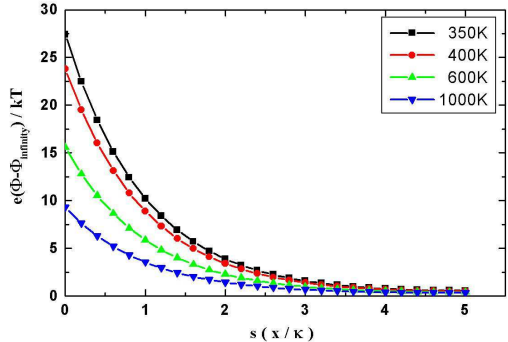}
\caption{\label{fig:SpChO6} Free surface space charge in YBa$_2$Cu$_3$O$_6$. $\Phi_{\infty}$ refers to the bulk
potential caused by space charge. S is the reduced length, scaled by the
Debye length. } \end{figure}

\noindent Boundary conditions for the slab of thickness 2L are given by

\begin{eqnarray} \rm
s=0, z= - {e(\Phi_\infty) \over kT}, \\
\rm
s={L \over \kappa}, z= {e(\Phi_L-\Phi_\infty) \over kT},
\end{eqnarray}

\noindent where $\Phi_L$ is the potential at x=L. In this study, we 
assume the thickness approaches infinity, which means the grain boundary 
is qualitatively modeled by two surfaces ``back to back". Under 
this condition $\Phi_L$ approaches $\Phi_\infty$. We solved the Eq.(\ref{eq:o6o6}) by 
conventional numerical integration techniques for several temperatures. 
The results are plotted in Fig.\ref{fig:SpChO6}. Since there is no 
oxygen at chain sites for $\rm YBa_2Cu_3O_6$, the chain copper is 
comparatively loosely bound inside crystal. Based on
the point defect energy calculations, it turns out that it costs the least
energy to make such a copper vacancy (see Table (\ref{tab:PointDefect})). 
Because the free energies of formation of the
anion and cation vacancies differ, there exists a charged surface and a
region of space charge of the opposite sign adjacent to the surface. For
$\rm YBa_2Cu_3O_6$, this leads to a positively charged
surface and a region of negative space charge beneath the surface. Presumably this also 
would be the case for dislocations and grain boundaries.

Based on the point defect energies in $\rm YBa_2Cu_3O_6$ (Table (\ref{tab:PointDefect})),
the Debye length $\kappa$ and
bulk potential $\Phi_{\infty}$ are tabulated in Table (\ref{tab:O6}) for
various temperatures. As the temperature rises, the concentrations of
point defects increase exponentially such that the Debye length
becomes reduced at higher temperature. Next, let's consider the trend of bulk potential
versus temperature. In the RHS of Equation (\ref{eq:poissonO6}), it is seen
that at lower temperature there is a bigger difference among
$\rm \exp(-{E_{ION} \over kT})$. To balance this, a larger absolute
value of the bulk potential relative to the surface is needed. Since this potential is positive, this
expels holes from the surface. In other words, the positively charged
surface can cause a serious depletion of the hole content there.

\subsubsection{Metallic Case: YBa$_2$Cu$_3$O$_7$}

Compared with YBa$_2$Cu$_3$O$_6$, the defect energy ``spectrum" as calculated with the shell model
with pair potential parameters from \cite{2baetzold88} is different for YBa$_2$Cu$_3$O$_7$
(see Table (\ref{tab:PointDefect})). The similar comparison has also been discussed by Baetzold
in studying point defects in YBa$_2$Cu$_3$O$_7$ and YBa$_2$Cu$_3$O$_{6.5}$ \cite{2baetzold88,2baetzold91}.
Since the plane copper is coordinated with four oxygens, it is quite 
stable in the sense that more energy is needed to make a vacancy at 
the copper site in CuO$_2$ planes. Interestingly, our atomistic 
simulations show that the formation energy of oxygen vacancies in the
CuO$_2$ plane is the lowest one in YBa$_2$Cu$_3$O$_7$. Another very important difference
is the existence of screening by mobile holes in YBa$_2$Cu$_3$O$_7$,
which is absent in YBa$_2$Cu$_3$O$_6$ because it is an insulator. In our
atomistic simulations of the formation energy of point defects we ignore the explicit screening of
defects by mobile holes. However, the screening by holes is included in
the space charge calculations, assuming holes are localized on ion sites (but mobile by
hopping between ions). Considering the temperature of fabrication of 
YBa$_2$Cu$_3$O$_7$ is much higher than T$_c$, YBa$_2$Cu$_3$O$_7$ is in 
fact in a normal state. The Ioffe-Refel resistivity of YBa$_2$Cu$_3$O$_7$ 
at normal state is very large, about 0.1 m$\Omega$cm, due to the small 
carrier concentration \cite{oren90,gun03}. Emery and Kivelson suggested 
this type of exotic ``bad metal" can be still regarded as a quasiparticle 
insulator \cite{emery95}. These observations encourage us to apply the classical 
approach to study the effects of band-bending on hole's segregation near 
surfaces at fairly high temperature (above 600K), which YBa$_2$Cu$_3$O$_7$ 
remains in metallic normal state.  

Following a procedure similar to that used for YBa$_2$Cu$_3$O$_6$, but now including chain oxygen vacancies
and localized (but mobile) holes, from $\rm \delta F = 0 $ the equilibrium concentrations are

\begin{equation} \rm
\left\{ \begin{array}{cc}
\rm    n_Y    =  N exp(- {E_Y-3e \Phi \over kT}), \\
\rm    n_{Ba} = 2N exp(- {E_{Ba}- 2e \Phi \over kT}), \\
\rm    n_{Cu_c} = N exp(- {E_{Cu_c}- 2e \Phi \over kT}), \\
\rm    n_{Cu_p} = 2N exp(- {E_{Cu_p}- 2e \Phi \over kT}), \\
\rm    n_{O_p} = 4N exp(- {E_{O_p} + 2e \Phi \over kT}), \\
\rm    n_{O_a} = 2N exp(- {E_{O_a} + 2e \Phi \over kT}), \\
\rm    n_{O_c} = N exp(- {E_{O_c} + 2e \Phi \over kT}), \\
    \end{array}
\right.
\end{equation}

\noindent where $\rm n_{O_c} $stands for the density (per unit volume) of vacancies at
chain oxygen sites. Since mobile holes are present in  YBa$_2$Cu$_3$O$_7$, the charge density is
given by

\begin{widetext}
\begin{equation} \rm
\rho(x) = e ( 2n_{O_p} + 2n_{O_a} + 2n_{O_c} - 3n_Y -2 n_{Ba} -
n_{Cu_c} - 2n_{Cu_p} + n_{hole}).
\end{equation}
\end{widetext}

To calculate the bulk potential $\Phi_\infty$, we assume that
holes are bound to oxygen ions in the $\rm CuO_2$ plane such that the
effective valence of oxygen changes from $-2$ to $-1.75$. Within this
approximation, we compute the individual defect energy by the
Mott-Littleton approach. Then by applying the charge neutrality constraint
for the bulk, the bulk potential $\Phi_\infty$ is obtained. The resulting values of
bulk potential and Debye length are tabulated in Table (\ref{tab:O7}).

\begin{table}
\begin{center}
\caption{Bulk Potential and Debye Length of
$\rm YBa_2Cu_3O_7$.}\label{tab:O7}
\begin{tabular}{rrr}
\hline
     T (K) &   ${e\Phi_{\infty}\over kT}$ &   $\kappa$ (m) \\ \hline
    600 &       3.55 &   1.58x10$^{-11}$ \\
    800 &       2.76 &   1.82x10$^{-11}$ \\
   1000 &       2.16 &   2.04x10$^{-11}$ \\ \hline
\end{tabular}
\end{center}
\end{table}

Notice that when including the screening effect caused by mobile holes,
the bulk potential is invariant. However, the Debye length changes
dramatically because of the hole concentration. At the optimally
doped case, there is around 0.2 hole per CuO$_2$ \cite{Zhang93}.
Therefore, the total hole content is about 0.4 per unit cell of $\rm
YBa_2Cu_3O_7$. We assume that holes are localized, but, can hop from
site to site. Hence, the hole content at the position of x with
the electrostatic $\Phi$(x) inside the slab is given as

\begin{equation} \rm
n_{hole}(x)    = N\cdot f_0 \times  \exp{- e (\Phi(x)-\Phi_{\infty})
\over kT }
\end{equation}

\noindent where $N \cdot f_0$ is the density of holes at surfaces,
determined by the following condition

\begin{equation} \rm
0.4 \times  N \times V = \int dA \int dx N \cdot f_0 \times \exp {- e
(\Phi-\Phi_{\infty}) \over kT };
\end{equation}

\noindent where $dA = dy*dz$ is the area. After the same process as 
described in the previous section, the Poisson equation yields

\begin{widetext}
\begin{eqnarray} \rm
\label{eq:o7o7}
{ (d^2z) \over (ds^2) } & = & \rm -{1 \over 2}[2 \times 4 \times
\exp(- {E_{O_p}+ 2e\Phi_\infty \over kT })\exp(-2z) \nonumber \\
\rm
& + & 2 \times 2 \times \exp(- {E_{O_a}+ 2e\Phi_\infty \over kT
})\exp(-2z)
+ 2\times 1\times \exp(- {E_{O_c}+2e\Phi_\infty \over kT })\exp(-2z)
\nonumber \\
\rm
& - & 3\times 1\times \exp(-{E_Y-3e \Phi_\infty \over kT})\exp(3z) -
2\times 2\times \exp(-{E_{Ba}-2e \Phi_\infty \over kT})\exp(2z)  \nonumber
\\
\rm
& - & 2\times 1\times \exp(-{E_{Cu_c}- 2e \Phi_\infty \over kT})\exp(2z)
 -2\times 2\times \exp(-{E_{Cu_p}-2e \Phi_\infty \over kT})\exp(2z)
 \nonumber \\
\rm
& + & f_0\times \exp(-z)].
\end{eqnarray}
\end{widetext}

\begin{figure}
\includegraphics [scale=0.4] {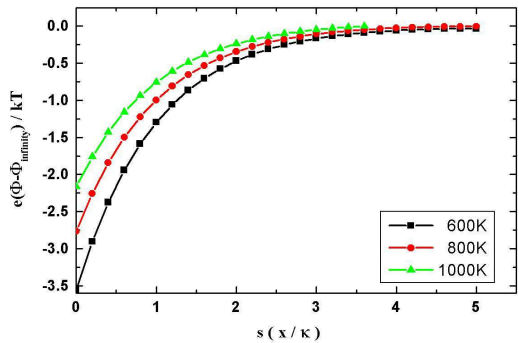}
\caption{\label{fig:SpChO7}
Free surface space charge in YBa$_2$Cu$_3$O$_7$. $\Phi_{\infty}$ refers to the bulk
potential caused by space charge. S is the reduced length, scaled by
the Debye length. }
\end{figure}

In YBa$_2$Cu$_3$O$_7$, based on the shell model calculations of point defect energies, it
turns out that it is much easier to create oxygen vacancies in the
CuO$_2$ plane than any other type of vacancy (See Table (\ref{tab:PointDefect})). The space charge
profile is obtained by numerical solutions of Eq.(\ref{eq:o7o7})
at several temperatures, which are plotted in Fig.\ref{fig:SpChO7}.
In our model, in contrast to $\rm YBa_2Cu_3O_6$ (insulator), $\rm YBa_2Cu_3O_7$ is a
p-type superconductor, containing mobile holes. Assume that
the total hole content is 0.4 per unit cell, this makes the dominant
contribution to the total charge density such that the total charge
density is effectively constant over the temperatures range of interest.
As for the bulk potential relative to the surface calculated by the procedure described in the
previous section its value is determined by differences among the $\rm \exp(-{E_{ION} \over kT})$ terms. 
So, at higher temperature, a much smaller bulk potential is needed to keep the bulk
neutral. As the total charge density is effectively constant, the Debye
length is proportional to the square root of the temperature [see Eq. (6)]. It is
intrinsically different from the corresponding trend for
YBa$_2$Cu$_3$O$_6$. Furthermore, note that the Debye length of O$_7$ is
far shorter than that of O$_6$, the mobile hole provides strong screening
of the charge at grain boundaries. We also computed the Thomas-Fermi
screening length $\rm \lambda_{TF}$ by the following relation
\cite{kittel6}:

\begin{equation} \rm
\lambda_{TF} = {\hbar\over e} ({\epsilon \over 4m_{hole}})^{1\over 2} 
({\pi
\over
3 \times (0.4\times N)})^{1\over 6},
\end{equation}

\noindent where $\rm m_{hole}$ is the effective mass of the hole, N is the number of YBa$_2$Cu$_3$O$_7$ unit cell per volume,
and we assume the total hole density equals to $\rm 0.4\times N$ for optimal doping. The
calculated $\rm \lambda_{TF}$ is around 1 $\AA$ by taking $\rm
m_{hole}= m_{ele}$. It is interesting to note that both lengths, the Thomas-Fermi screening length $\rm \lambda_{TF}$ 
and the Debye length $\kappa$,
are shorter that an interionic spacing. Because holes are localized, the
effective mobile hole density is expected to be only
small amount of the total density. In fact, the $\kappa$ and $\rm
\lambda_{TF}$ are the lower limit of the effective screening length.
The very short screening length can help to protect
superconducting properties inside bulk from the
influence of grain boundaries if the hole density of grain boundaries
corresponds to YBa$_2$Cu$_3$O$_7$. However, experiments have shown that
the oxygen content
at grain boundaries is often less than that inside the bulk [2]. If oxygen
content is less than around O$_{6.45}$, from the calculation in the
previous section it can be concluded that the
Debye length will increase dramatically. It means that a large portion of
the bulk near the boundary will be affected by the internal electric field
caused the space charge around grain boundaries. This may be one reason
for the vicinity of some grain boundaries not being a good superconductor.

\subsection{Space charge of calcium-doped YBa$_2$Cu$_3$O$_7$}

After the discovery of YBa$_2$Cu$_3$O$_7$, many types of ions have been
chosen to dope into YBa$_2$Cu$_3$O$_7$ in search of new superconductors and
to improve properties. Doping with Ca into [001] tilt grain boundaries
of YBCO has been shown to increase the critical current
density above $10^5 (A / cm^{2})$ \cite{Hammerl00}. Our goal here is to study
the segregation of calcium around small angle tilt boundaries of Y-123 in order to aid in 
the search for a more
efficient approach for doping Ca. First, doping calcium in bulk is
studied by shell model calculations to identify the energetically favorable reaction for doping. 
Secondly, the volume of solution of a Ca dopant atom is calculated.
Thirdly, the segregration of calcium is represented in terms of the
segregation energy as function of angle around the core of edge dislocations in tilt boundaries, which can
clarify the role of the elastic strain field. Finally, the space charge profile is studied by including extra holes due 
to dopants and segregation of dopants due to the strain field.

\subsubsection{Site Preference}

When Ca replaces Y, it is accompanied by a hole for the sake of charge
neutrality. However, when substituting Ca for Ba, the charge remains
balanced without changing the total hole content. Thus, it is important to
know the circumstances when Ca doping yields Ca on Y or Ca on Ba sites. Here, we
mainly focus on the dissolution of divalent
Ca$^{2+}$, which can be represented by the following defect
reactions, depending on which sublattice Ca$^{2+}$ chooses and what kind
of charge compensation is involved. The parameter m is chosen to
characterize the dopant concentration in the YBCO.

\begin{widetext}
\begin{eqnarray}
\label{eq:homo}  \rm
CaO & + & \rm m YBa_2Cu_3O_7  \\ \nonumber
\rm
& \rightarrow  & m Y(Ba_{2m-1 \over 2m}Ca_{1\over 2m})_2Cu_3O_7 + BaO ; \\
\label{eq:chain}
\rm
2CaO & + & 2m YBa_2Cu_3O_7 + { 1\over 2} O_2 \\ \nonumber
\rm
& \rightarrow & 2m(Y_{m-1 \over m}Ca_{1\over m})(BaO)_2(Cu^{3+}O)_{1\over
2m}(CuO)_{2m-1\over 2m}(CuO_2)_2 + Y_2O_3 ; \\
\label{eq:plane}
\rm
2CaO & + & 2m YBa_2Cu_3O_7 + {1\over 2}O2 \\ \nonumber
\rm
& \rightarrow & 2m (Y_{m-1 \over m}Ca_{1\over
m})(BaO)_2(CuO)(Cu^{3+}O_2)_{1\over2m}(CuO_2)_{4m-1 \over 2m} + Y_2O_3;
\end{eqnarray}
\end{widetext}

The first case, Eq.(\ref{eq:homo}), is homovalent substitution. Simply,
barium is replaced by calcium. However, yttrium is replaced in the last 
two reactions (Eqs.(\ref{eq:chain}) and (\ref{eq:plane})). These two 
reactions are aliovalent substitutions, in which it is very important to 
identify the mechanism of charge
compensation. In reaction (\ref{eq:chain}), a hole is created in the CuO
chain to make the whole system charge neutral. Assuming the hole is
trapped at copper site, the valence of one copper at chain site increases
by one approximately. In reaction (\ref{eq:plane}), we assume that the
extra hole is in the CuO$_2$ plane, such that the valence of one copper in
the plane increases by one. All the calculations are done by assuming the
infinite dilution of dopants. When the valence of copper changes, the
corresponding interatomic pair potentials are scaled based on a
simple radius scheme \cite{hbsu-unpub}. GULP was used to compute the
defect energies. The calculated energies of solution (per
calcium) are listed in Table (\ref{tab:Cahole}). Previous studies on bipolaron binding energies have shown that the hole is more likely to
localized on copper sites than oxygen sites in the CuO$_2$ planes \cite{2baetzold88}. In our studies, the hole content is tuned by
the dopant, calcium. The calculations suggest that it is most energetically favorable to substitute a calcium ion at
an yttrium site, together with the creation one hole in the CuO$_2$ plane. This agrees very
well with experimental measurements \cite{Hammerl00}.

\begin{table}
\begin{center}
\caption[Energies of Solution of Calcium at Infinite Dilution
in YBa$_2$Cu$_3$O$_7$ .]
{Calculated Energies of Solution of Calcium at Infinite Dilution in
YBa$_2$Cu$_3$O$_7$ (in eV per Ca$^{2+}$)}
\label{tab:Cahole}
\begin{tabular}{rr}
\hline
Reaction (\ref{eq:homo})  &       3.14 \\
Reaction (\ref{eq:chain}) &       2.99 \\
Reaction (\ref{eq:plane}) &       1.02 \\ \hline
\end{tabular}
\end{center}
\end{table}
\subsubsection{Volume of Solution for doping AE(2+) at Y(3+) Site}

Doping Ca in grain boundaries of YBCO can improve the transport properties
significantly since more holes are created near grain boundaries. The
strong elastic strain field around grain boundaries has an important role
in the segregation of dopants. The disruption of the lattice by a point
defect, such as a dopant atom like Ca$^{2+}$ substituting for Y$^{3+}$ in
YBCO, gives rise to forces on the surrounding lattice atoms which, in
turn, give rise to a displacement field in the lattice around the defect.
The interaction between the defect displacement field and the stress
fields of other defects, such as dislocations, or applied stress fields,
alter the energy of the defect \cite{Teoclosiu82,girifalco67,eshelby56} 
and cause defect segregation. The effects of
the forces on the lattice due to the point defect can be described by
means of force multipoles \cite{Teoclosiu82}, which together
with the Green's tensor function of the matrix of the perfect crystal, can
be used to calculate the resulting displacement field caused by the defect
in the crystal. In this section, we use numerical shell model calculations
with GULP to estimate the strength of the elastic force dipole for the
point defects (dopants). Then in the next section we apply anisotropic
continuum elasticity to describe the stress field due to an edge
dislocation; finally we use continuum elasticity theory to compute the
elastic interaction between the dopant atom and the dislocation. Such
methods have been shown to give good agreement with direct numerical
calculations of the interaction between external pressure and vacancies in
simple crystals such as van der Waals bonded rare gas crystals (see
appendix 3 in reference \cite{girifalco67}).

To study the segregation of calcium around tilt grain boundaries, it is
necessary to have the volume of solution to compute the contribution to
the binding energy of Ca$^{2+}$ at grain boundaries due to interactions
with the strain field of edge dislocations in the boundaries. In the
force-dipole approximation, the difference between the energy to
form the defect in a crystal containing a strain field $\epsilon_{ij}$
(either from an applied stress or from the stress field of another
defect) and the energy in the strain-free crystal is given as

\begin{equation}
\label{eq:Eint}
E_{int} = - a_{ij} \epsilon_{ij},
\end{equation}

\noindent where the Einstein summation convention for repeated indices is
used and the coefficients a$_{ij}$ characterize the strength of the defect
forces on the lattice. The effective body force, f, which the defect
exerts on the lattice, is:

\begin{equation}
f_i = - a_{ij} {\partial \over \partial x_j} \delta (r),
\end{equation}

\noindent where $\delta(r)$ is the Dirac delta function. These three force
components represent three orthogonal ``double forces without moment" and
are sometimes called ``force dipoles". The values of a$_{ij}$, which have
the unit of energy, must be obtained from experiment or numerical
simulations. For a defect with spherical symmetry in an elastically
isotropic matrix:

\begin{equation}
a_{ij} = a \delta_{ij},
\end{equation}

\noindent where $\delta_{ij}$ is the Kronecker $\delta$ (=1, if i=j; =0,
if i$\neq$j), and

\begin{equation}
a = B \Omega,
\end{equation}

\noindent where $\Omega$ is the volume change produced by the formation of
the defect in a finite crystal and B is the bulk modulus \cite{eshelby56}.
Thus from equation (\ref{eq:Eint}) we obtain

\begin{equation}
E_{int} = - B \Omega (\epsilon_{11} + \epsilon_{22} + \epsilon_{33}) =  P
\Omega ,
\end{equation}

\noindent where P is the hydrostatic component of pressure at the defect,
produced by external forces or other defects.

In the case of a defect with orthorhombic symmetry, equation
(\ref{eq:Eint}) yields

\begin{equation}
E_{int} = - [a_{11}\epsilon_{11} + a_{22}\epsilon_{22} +
a_{33}\epsilon_{33} ].
\end{equation}

\noindent Clearly, it is necessary to evaluate E$_{int}$ for three
uniaxial strain cases. For strain along a-axis ($\epsilon_{11} \neq 0;
\epsilon_{22}=\epsilon_{33}=0$), we have

\begin{equation}
a_{11} = - {E_{int} \over \epsilon_{11}},
\end{equation}

\noindent where E$_{int}$ is evaluated from numerical simulation by GULP.
Using generalized Hooke's law in the form of elastic compliances, S$_{ij}$,
we obtain by choosing the crystal axes as principle ones

\begin{eqnarray}
\epsilon_{11} & = & S_{11}\sigma_{11} + S_{12}\sigma_{22} +
S_{13}\sigma_{33}, \\ \nonumber
\epsilon_{22} & = & S_{12}\sigma_{11} + S_{22}\sigma_{22} +
S_{23}\sigma_{33}, \\ \nonumber
\epsilon_{33} & = & S_{13}\sigma_{11} + S_{23}\sigma_{22} +
S_{33}\sigma_{33} .
\end{eqnarray}

\noindent Hence, the components of volume of formation of dopants are
given as

\begin{eqnarray}
\Omega_{11} & = & S_{11}a_{11} + S_{12}a_{22} +
S_{13}a_{33}, \\ \nonumber
\Omega_{22} & = & S_{12}a_{11} + S_{22}a_{22} +
S_{23}a_{33}, \\ \nonumber
\Omega_{33} & = & S_{13}a_{11} + S_{23}a_{22} +
S_{33}a_{33} .
\end{eqnarray}

\begin{figure}
\includegraphics [scale=0.4] {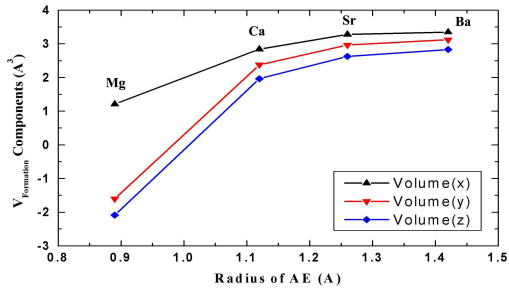}
\caption{\label{fig:CaVol}
Components of Volume of Formation for doping AE$^{2+}$ at a Y$^{3+}$
site of YBa$_2$Cu$_3$O$_7$ vs. Radius of AE. }
\end{figure}

The calculated components of the volume of solution for various divalent ions (AE$^{2+}$) at
the Y$^{3+}$ site of
YBa$_2$Cu$_3$O$_7$ are plotted in Fig.\ref{fig:CaVol}. Considering the
trend across the series of AE$^{2+}$, the ion size plays the important
role in the stress effects caused by doping AE$^{2+}$ at Y$^{3+}$. The
point defect for doping AE$^{2+}$ at Y$^{3+}$ site carries one negative
charge. This strongly repels the surrounding oxygen ions to make the sign
of the volume of solution positive for AE ions whose radius is similar
or larger than that of Y$^{3+}$ (1.02 $\AA$). Note that the short range repulsion
between AE and surrounding oxygen ions becomes larger with increasing AE
ion radius, it is expected that the volume of solution increases with AE
ion radius. However, for an AE ion with a larger radius, the polarization of the lattice around the solute
which helps to lower defect's energy by polarizing electron ``clouds"
also depends on the distortion of the lattice by the AE ion and thus contributes to the
volume of formation. Therefore, because of these competing effects the volume of formation ends up
with a ``satuation-like" curve.


\subsubsection{Dislocation Stress Fields in Y-123}

Grain boundaries of importance in coated conductors are
low-angle [001] tilt boundaries. Thus they can be considered as an array
of isolated edge dislocations. We observe that the elasticity in YBa$_2$Cu$_3$O$_7$ 
is not far different between the a-
and b-axis ( (C$_{xx}$-C$_{yy}$)/C$_{yy}$ $\sim$ 0.082), so this orthorhombic crystal can be approximated by a hexagonal 
one, which is motivated by the existence of closed-form expressions for
the stress fields in the case of anisotropic elasticity in hexagonal
crystals. Here we average elastic constants of orthorhombic symmetry into those of
permitted by hexagonal symmetry by taking arithmetical average of
corresponding elastic moduli.
The geometry of the dislocations in [001] tilt boundaries in Y-123 is like
prismatic slip in hexagonal crystals. The Burger's vector
lies in basal plane (${a\over 3}<1\bar{2}10>$), while dislocation line
runs along the hexad axis. The prismatic planes are $<{\bar{1}010}>$.
Following Ref. \cite{Steeds73}, on account of the basal plane
isotropy, the stress field is given by

\begin{eqnarray}
\sigma_{xx} & = & {K_1 \cdot b \cdot y \cdot (3x^2 + y^2) \over
2\pi \cdot (x^2+y^2)^2} , \\
\sigma_{yy} & = & { - K_1 \cdot b \cdot y \cdot (x^2 - y^2) \over
2\pi \cdot (x^2+y^2)^2} , \\
\sigma_{zz} & = & - {s_{xz} \over s_{zz}}\cdot
(\sigma_{xx}+\sigma_{yy});
\end{eqnarray}

\noindent where s is the elastic compliance modulus, and $K_1 = {s_{zz}\over
[2\cdot s_{xx}s_{zz}-s_{xz}^2]}$. (Note: The formula of $K_1$ on page 192
of Ref. \cite{Steeds73} has one misprint.)

Small angle (usually less than 5$^o$) grain boundaries can be considered 
as an array of isolated dislocations. The separation between dislocations 
is around $d \cong {b\over \theta} $. If we assume that a superposition 
approximation can be applied here, the total stress field can be written 
as

\begin{equation}
\sigma_{ii}^{total} = \sum_{n=-2,..,+2} \sigma_{ii}(y+nd).
\end{equation}

When doping calcium at yttrium sites, the non-zero tensional
elements of the ``volume of solution" will interact with the stress field.
Neglecting the effect of the solute atom on the local elastic modulii,
the segregation energy of calcium can be written as

\begin{equation}
E_{seg} = \sum_{i=x,y,z} \sigma_{ii}^{total} \Omega_{ii}.
\end{equation}

\begin{figure}
\includegraphics [scale=0.5] {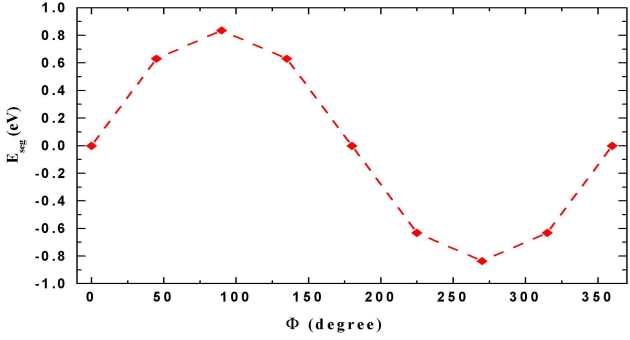}
\caption{\label{fig:CaSeg}
Elastic components of the segregation energy of calcium around the center of an edge dislocations in small angle tilt grain
boundaries YBa$_2$Cu$_3$O$_7$. $\Phi$ is the angle starting from (0,b),
rotating counterclockwise around the core of small angle (usually less 
than 5$^o$) tilt grain boundaries for a
distance of one Burgers vectors from the center.}
\end{figure}

In the previous section, we calculated the energy for doping calcium
at yttrium sites inside a stress free bulk Y-123. Here we computed the
segregation energy for calcium due to the stress field around very small 
angle (usually less than 5$^o$) tilt boundaries, which is plotted in 
Fig.\ref{fig:CaSeg}. Since our region of interest is at a small distance 
from the dislocation cores, there exist possible errors in using linear 
elasticity theory here and our results are probably overestimated. 
However, the relative energy differences are more important than absolute 
values. In Fig.\ref{fig:CaSeg}, the calculations show that calcium 
segregation is expected to the tensile stress region due to the
positive sign of volume of formation. This will
affect the distribution of mobile hole around grain boundaries. We will
study this further in the following section for general grain boundaries.


\subsubsection{Effects on space charge due to segregation of calcium near grain boundaries}

It has been noted \cite{Eshelby58} that edge dislocations could act as
sources and sinks for vacancies such that space charges would appear also in the regions near
dislocation cores. The modeling of space charges around dislocation cores is
more difficult than our treatment of the intrinsic space charges at free surfaces from the difference in formation energy
of equilibrium point defects. If for no other reason than that different 
in the energy required for creation of point
defects by removing ions from the bulk and placing them in the dislocation 
core,it is not likely to be quite the same as that for
free surfaces. However there is also likely to be an alteration of the 
space charge by dopants and their accompanying holes if
the dopants tend to segregate to the dislocations. In the previous section (II B.3) it was shown that there is a strong 
attraction
of the AE$^{2+}$ dopants, i.e. Ca, Sr, and Ba, to the tensile regions of isolated edge dislocations in low angle grain 
boundaries
and a repulsion from the compressive regions. We will illustrate some of the effects of such dopant segregation on space 
charges by an
idealized planar grain boundary which consists of joining together two free surfaces with their associated point defect space
charges, together with Ca$^{2+}$ dopants and accompanying holes which have segregated in the stress fields of the boundary. 
We assume that the dopant atoms are attracted to or repelled from region along the boundary by the elastic fields 
which decay going away from the boundary. The results of this calculation will
illustrate the interplay between dopant segregation and point defect induced by space charges.

From previous thermodynamic calculations we know that
hole density changes with the content of doped calcium.
Consequently, the Debye length varies with Ca concentration.
Note that the defect created by replacing Y$^{3+}$ with Ca$^{2+}$ carries one
negative charge. Assuming the total number of Ca$^{2+}$ is conserved,
after solving the distribution of charged particles in an electrostatic
field caused by space charges, the defect density $\rm n_{Ca}$ at the position
of x with the electrostatic $\Phi$(x) away from the boundary plane is given as

\begin{equation} \rm
\label{eq:ca}
n_{Ca} = N \cdot f_{Ca} *  \exp{ e (\Phi-\Phi_{\infty}) \over kT
} ;
\end{equation}

\noindent where $N \cdot f_{Ca}$ is the density of calciums at surfaces
determined by the following condition

\begin{equation} \rm
N\cdot n_{doping}^{Ca}\cdot V = \int dA \int dx N \cdot f_{Ca}
\cdot \exp({ e (\Phi-\Phi_{\infty}) \over kT }),
\end{equation}

\noindent where $\rm n_{doping}^{Ca}$ is the total concentration of doped
calciums. Since each Ca$^{2+}$ is accompanied by one mobile hole, the
density of holes near surfaces ($N \cdot f_0$) is determined by the
following condition (assuming the total number of holes are conserved)

\begin{equation} \rm
(0.4 + n_{doping}^{Ca}) N\times V = \int dA \int
dx N \cdot f_0
\cdot \exp {- e (\Phi - \Phi_{\infty}) \over kT }.
\end{equation}

Note that we extend our studies to general grain boundaries here based on 
understanding divalent dopants' interaction with stress field of small 
angle tilt boundaries. As there can be a significant strain field 
in the grain boundaries region, originating from size mismatch between 
solute and matrix atoms,
dislocation, etc, a strain energy term should be included in
the studies of Ca-doping because the strain field due to
dislocations at grain boundaries can substantially affect the segregation
of Ca. We ignore the elastic interactions with the boundaries of the charged vacancies,
and approximate the effect of the stress field of the boundary by an elastic term of the calcium by
the same form used in Ref. \cite{Yan83a}:

\begin{eqnarray} \rm
U_{elastic} &=& \rm U_0 (1-({s\cdot \kappa \over d})^n),   s \cdot \kappa \leq 2d;
\nonumber \\
&=& 0,      \hspace{0.1in}        \rm     s \cdot \kappa > 2d .
\end{eqnarray}

\noindent where U$_0$ is the binding energy near the dislocation core. We
choose n=2, and d is about two lattice parameters
along b-axis. This is consistent with the periodicity of the structural
units proposed in the coincidence models of grain boundary structures
\cite{gleiter72}. Calculations in the previous section indicated that the volume
of solution for doping calcium at yttrium sites is positive. Hence, the
positive sign of $\rm U_0$ means in the compression
region which repels Ca, while the negative sign of $\rm U_0$ means in the tensile region which attracts Ca.
The charge density now is given by

\begin{widetext}
\begin{equation} \rm
\rho(x) = e ( 2n_{O_p} + 2n_{O_a} + 2n_{O_c} - 3n_Y -2 n_{Ba} -
n_{Cu_c} - 2n_{Cu_p} + n_{hole} - n_{Ca}).
\end{equation}
\end{widetext}

Following the same process as previous section, the Poisson equation can be
re-written as

\begin{widetext}
\begin{eqnarray} \rm
{ (d^2z) \over (ds^2) } & = & \rm -{1 \over 2 }[2\times 4\times
\exp(- {E_{O_p}+ 2e\Phi_\infty \over kT })\exp(-2z) \nonumber \\
& + & 2\times 2 \times \exp(- {E_{O_a}+ 2e\Phi_\infty \over kT })\exp(-2z)
+ 2\times 1\times \exp(- {E_{O_c}+2e\Phi_\infty \over kT })\exp(-2z)
\nonumber \\
& - & 3\times 1\times \exp(-{E_Y-3e \Phi_\infty \over kT})\exp(3z) -
2\times 2\times \exp(-{E_{Ba}-2e \Phi_\infty \over kT})\exp(2z)  \nonumber
\\
& - & 2\times 1\times \exp(-{E_{Cu_c}- 2e \Phi_\infty \over kT})\exp(2z)
 -2\times 2\times \exp(-{E_{Cu_p}-2e \Phi_\infty \over kT})\exp(2z)
 \nonumber \\
& + & f_0\times \exp(-z) - f_{Ca} \times \exp(z - {U_{elastic}\over kT}) ].
\end{eqnarray}
\end{widetext}

When calcium is doped in YBCO, the process is often designed so that
calcium diffuses into grain boundaries first. However, it is difficult
to estimate the amount of calcium at the grain boundaries during the
deposition process. In this section, we assume the hole density and
the amount of calcium near grain boundaries are 0.6 and 0.08 per unit
volume, respectively. Furthermore, the elastic binding energy of calcium
is chosen to include both compression and tension regions: $U_0 = (-6, -4,
-2, 0, 2, 4, 6)kT$. This is shown in Fig.5 for an idealized boundary in 
YBa$_2$Cu$_3$O$_7$. As discussed in previous
section, the grain boundaries in O$_7$ may be negatively charged because
of the intrinsic space charges due to point defects. It should
be also stressed that calcium is doped at yttrium sites, such that this
site has one negative charge accompanied by making holes at CuO$_2$
planes. Therefore, the screening length becomes even shorter than that
of pure $\rm YBa_2Cu_3O_7$. If the calcium is pushed away from grain
boundaries, the space charge remains almost the same (somehow little
dependent on repulsive energy). However, on the other hand, if the calcium
is attracted by grain boundaries, extra negative charge will accumulate
around grain boundaries such that the space charge's profile is
significantly adjusted. This can strongly perturb the distribution of
mobile hole density around grain boundaries. The stress field around grain
boundaries is also so strong (around $10^2$ GPa based on our calculated
elastic properties of YBCO that it significantly
affects the surrounding oxygen content and ordering \cite{hbsu-unpub}). If oxygen content
is close to O$_6$, the grain boundaries are possibly positively charged, and the
large Debye length leads to a larger portion of the bulk being exposed
into the influence of grain boundaries. Under such circumstances, doping
with calcium not only neutralizes (by the intrinsic negative charge of
Ca$_Y^{-1}$) the charged boundaries, but also increases the density of mobile
holes to screen the charged boundaries. Therefore, the bad low angle tilt boundaries are ``cured" by doping calcium.

\begin{figure}
\includegraphics [scale=0.4] {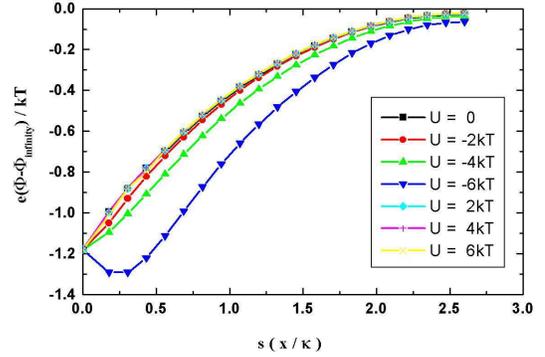}
\caption{\label{fig:SpChCa} Space charge of doping calcium at idealized $\rm
YBa_2Cu_3O_7$ grain boundaries at 723$^o$C. S is the reduced length, scaled by the
Debye length. U is energy due to strain field. Negative sign means attractive
interaction. }
\end{figure}

\section{Conclusions}

We have systematically studied the space charge profile for
YBa$_2$Cu$_3$O$_6$, YBa$_2$Cu$_3$O$_7$, and Ca-doped cases. Intrinsic
space charge is strongly dependent on the oxygen content of
YBa$_2$Cu$_3$O$_{7-\delta}$, which is determined by the annealing
temperature and oxygen partial pressure. For YBa$_2$Cu$_3$O$_6$
the chain copper is loosely bound with neighboring two oxygens, therefore
it is comparatively easier to create vacancies at copper chain sites for O$_6$.
Hence, the grain boundaries are possibly positively charged. On the other
hand, the calculations show that it requires less energy to make
vacancies at oxygen sites for O$_7$. Consequently, it was seen that there exists a
dramatic change in the nature of the space charge in going from O$_6$ to
O$_7$. It is interesting to note that the origin of the
screening in going from O$_6$ to O$_7$
is completely different. For O$_6$ it is the thermal activated vacancies,
while for O$_7$ it is the mobile holes. This is very similar to the two
screening mechanisms in semiconductors: one is intrinsic, the other is
free carriers due to doped impurities. In addition, We have carried out studies of the segregation of calcium at tilt 
grain boundaries in
the Y-123 compound. By comparing calculated heat of solution for various doping mechanisms, we have
found that the most energetically favorable defect reaction is that of doping calcium at yttrium sites
accompanied by creating one hole in the CuO$_2$ plane per Ca ion. This
conclusion is valid for the reaction in the bulk, which agrees with previous experimental
measurements \cite{Hammerl00}. To study the segregation of calcium at tilt boundaries, the
volume of formation is needed in order to estimate the binding energy to
the elastic strain field which arises from the grain boundary
dislocations. We computed this for the doping series of alkaline
earths at yittrium sites. After including the strain field contribution,
the segregation energy of calcium at small angle tilt boundaries is obtained. The
theoretical results suggest that more calcium is expected in the tensile
regions around tilt grain boundaries. For calcium-doped
YBa$_2$Cu$_3$O$_{7-\delta}$ case, our calculations indicate that Y$^{3+}$ ions are replaced by Ca$^{2+}$ ions,
accompanied by creating holes in CuO$_2$ plane. The increase of hole
content strongly enhances the screening effect by mobile holes. The strong
screening effect (much shorter screening length) helps to decrease the
area of influence of the grain boundaries. Besides, the
segregation of calcium is strongly determined by the strain field. More
calcium segregation is expected in tensile stress regions. Since the
calcium is doped at yttrium sites, it has one negative charge.
Therefore, the segregation of calcium causes extra negative defects to approach grain boundaries such
that the potential decreases quite significantly. Consequently, the
hole content near grain boundaries increases, thus helping to screening out the effect of disorder at the boundary.
In summary, the nature of space charge is closely related to the oxygen content. Doping
calcium at small angle tilt grain boundaries of YBCO leads to the increase of J$_c$
through increasing the hole content, enhancing negative potential regions due to the segregation of calcium, and 
helping to passivate disorder at the boundary.

{\em Note Added---} After submission of this manuscript, ref.\cite{zhu04} 
reported the experimental measurement of potential curves near [001] tilt 
boundaries for both YBa$_2$Cu$_3$O$_{7-\delta}$ and Ca-doped 
YBa$_2$Cu$_3$O$_{7-\delta}$. In particular, the curves in Fig.2 of 
ref.\cite{zhu04} were nicely predicted by Figs. 2 and 5 in this 
manuscript. (Indeed, the main body of this manuscript has been already 
completed as chapter 6 of ref.\cite{su-thesis}. The Figs. 6.2 and 6.3 of 
ref.\cite{su-thesis} are in fact plotted here as Figs. 2 and 5.)

\begin{acknowledgments}
HBS is grateful for Chinatsu Maeda's kind assistance in preparing the
manuscript. Work at Brookhaven National Laboratory was performed under 
auspices of the Division of Materials Sciences, Office of Science,
U.S. Department of Energy under contract No. DE-AC-02-98CH10886.
\end{acknowledgments}

\end{document}